\documentclass[graybox,natbib,round]{svmult}

\usepackage{type1cm}        
\usepackage{makeidx}         
\usepackage{graphicx}        
\usepackage{multicol}        
\usepackage[bottom]{footmisc}

\usepackage{newtxtext}       %
\usepackage{newtxmath}       
\usepackage{hyperref}

\makeindex             

\usepackage{xspace}

\newcommand{\msun}{\ensuremath{M_\odot}}
\newcommand{\nickel}{\ensuremath{^{56}}Ni\xspace}
\newcommand{\eexp}{\ensuremath{E_{\rm exp}}}
\newcommand{\mni}{\ensuremath{M_{\rm Ni}}}
\newcommand{\mej}{\ensuremath{M_{\rm ej}}}

\begin{document}

\title*{The explosion mechanism of core-collapse supernovae and its observational signatures}
\author{Ond\v{r}ej Pejcha}
\institute{Ond\v{r}ej Pejcha \at Institute of Theoretical Physics, Faculty of Mathematics and Physics, Charles University, V~Hole\v{s}ovi\v{c}k\'{a}ch 2, 180 00, Praha 8, Czech Republic, \email{pejcha@utf.mff.cuni.cz}}
%
%
\maketitle

\abstract{The death of massive stars is shrouded in many mysteries. One of them is the mechanism that overturns the collapse of the degenerate iron core into an explosion, a process that determines the supernova explosion energy, properties of the surviving compact remnant, and the nucleosynthetic yields. The number of core-collapse supernova observations has been growing with an accelerating pace thanks to modern time-domain astronomical surveys and new tests of the explosion mechanism are becoming possible. We review predictions of parameterized supernova explosion models and compare them with explosion properties inferred from observed light curves, spectra, and neutron star masses.\newline\indent}

\section{Introduction}
\label{sec:intro}

Massive stars develop iron or oxygen-neon-magnesium cores, which eventually experience instability and collapse to reach nuclear densities. The initial stellar mass separating white dwarf or neutron star formation is commonly placed at around $8\,\msun$ \citep[e.g.][]{smartt09}, but depends on metallicity and other parameters \citep[e.g.][]{ibeling13}. Stars with helium core masses $\gtrsim 30\,\msun$ are destabilized by creation of electron--positron pairs, which reduces their mass in one or more mass ejection episodes and which can even completely disrupt the star \citep[e.g.][]{woosley17}. The collapse of the inner regions of the core stabilizes when the repulsive part of the strong nuclear force becomes important and the equation of state stiffens. The core overshoots this new equilibrium, bounces, and forms a shock wave propagating outward from a nascent proto-neutron star. The outgoing shock stalls into an accretion shock due to energy losses from neutrino emission and photodissociation of infalling nuclei. A prolonged period of accretion ensues and lasts many dynamical timescales of the system, $\gtrsim 100$\,ms. During this phase, the proto-neutron star grows in mass and the region below the accretion shock becomes unstable to turbulence and standing-accretion shock instability \citep[e.g.][]{herant94,burrows95,blondin03}. The instabilities are partially driven by neutrinos emanating from the proto-neutron star and the accretion region. This complex system of instabilities in a region semi-transparent to neutrinos presents a great challenge to theoretical understanding.

The evolution of stalled accretion shock likely bifurcates into two possible outcomes. The accretion can continue until the central object collapses into a black hole with much of the rest of the star following its fate. This evolutionary path can be accompanied by a transient brightening fainter than a typical supernova \citep{lovegrove13}. In a majority of stars, however, it is believed that the combined action of neutrinos and instabilities overturns the accretion into explosion \citep[e.g.][]{janka16,burrows18}. It is worth noting that the concept of neutrino mechanism assisted by instabilities is far from proven. There are other less-explored paths to explosion involving magneto-rotational processes \citep[e.g.][]{leblanc70,burrows07} and energy transfer by waves \citep[e.g.][]{burrows06,gossan19}

In that case, the accretion shock starts traveling out in radius and eventually reaches the stellar surface. Since interactions of neutrinos with matter are important, it is illuminating to frame the bifurcation between a failed and a successful explosion within the context of critical neutrino luminosity required to overturn the accretion into explosion \citep[e.g.][]{burrows93,yamasaki05,murphy08}. This framework enables assessment of the impact of various physical processes on the explosion and provides a physically-motivated ``antesonic'' explosion condition \citep{pejcha12,raives18}. There are other, more or less related conditions based on outward acceleration \citep{bethe85}, timescales of heating and advection \citep[e.g.][]{thompson00,thompson05,murphy08,muller15}, and the dynamics of the net neutrino heating region \citep{janka01}. 

After the accretion shuts off, there can be a period of simultaneous accretion and outflow from the proto-neutron star. Eventually, the proto-neutron star develops a wind driven by absorption of neutrinos. These processes set the baryonic mass of the remnant neutron star. The shock propagating through the star heats up the stellar interior. Gas heated above $\approx 5\times 10^9$\,K undergoes nuclear burning to iron-group elements. After the shock leaves the surface of the star, we observe the hot and expanding ejecta as a core-collapse supernova. Part of the light output of the supernova comes from the radioactive decay of the newly synthesized elements, especially $^{56}$Ni. The asymptotic energy of the supernova ejecta, about $10^{51}$ ergs, is $\sim 1\%$ of the neutron star binding energy, and the energy radiated in the optical and nearby wavelengths is $\sim 1\%$ of the asymptotic ejecta energy. Ultimately, all of the supernova explosion energy is radiated away as the ejecta decelerates and mixes with the interstellar medium.

\subsection{Scope of this review}
When and how does the collapse of the stellar core turn to explosion has been a major unsolved problem in theoretical astrophysics. There have been wide-ranging efforts in both theory and multi-messenger observations to make progress on this issue. An eloquent summary of much of this work is available in a recently published Handbook of Supernovae \citep{hsn}, especially the chapters of \citet{hsn_janka,hsn_foglizzo,hsn_oconnor,hsn_limongi,hsn_zampieri,hsn_sim,hsn_jerkstrand,hsn_galyam,hsn_arcavi,hsn_nugent,hsn_podsiadlowski}. We will not repeat the covered subjects here. We suggest the reader to consult this resource for more details and more complete lists of references. 

In this review, we focus on a small niche of the core-collapse supernova problem: how do we compare predictions of supernova theory with observations in the era of massive time-domain surveys? For example, the \emph{Large Synoptic Survey Telescope} (LSST) will begin discovering $\sim 10^5$ core-collapse supernovae every year starting in 2023. It is tempting to utilize this wealth of data to learn about the explosion mechanism. However, the benefit of this rapidly expanding dataset is not immediately obvious, because the majority of the expected supernova discoveries will provide only limited information for each supernova: a sparsely sampled light curve and an occasional low-resolution classification spectrum. In fact, even nowadays the majority of transient discoveries remain spectroscopically unclassified. This type of data naturally leads to a shift of attention from individual objects to population studies, which presents new challenges. Supernova theory has long been driven by very computationally intensive simulations, which could be run only for a  limited sample of initial models. Such works have been revealing fascinating complications of the explosion mechanism, but their predictive power for populations of stars has been limited. In the past few years, however, theoretical and observational efforts have been converging to a point, where mutual comparison is possible. This interaction of observations with theory is what we aim to capture in this review.

\section{Theoretical predictions}

\begin{figure*}
    \centering
    \includegraphics[width=\textwidth]{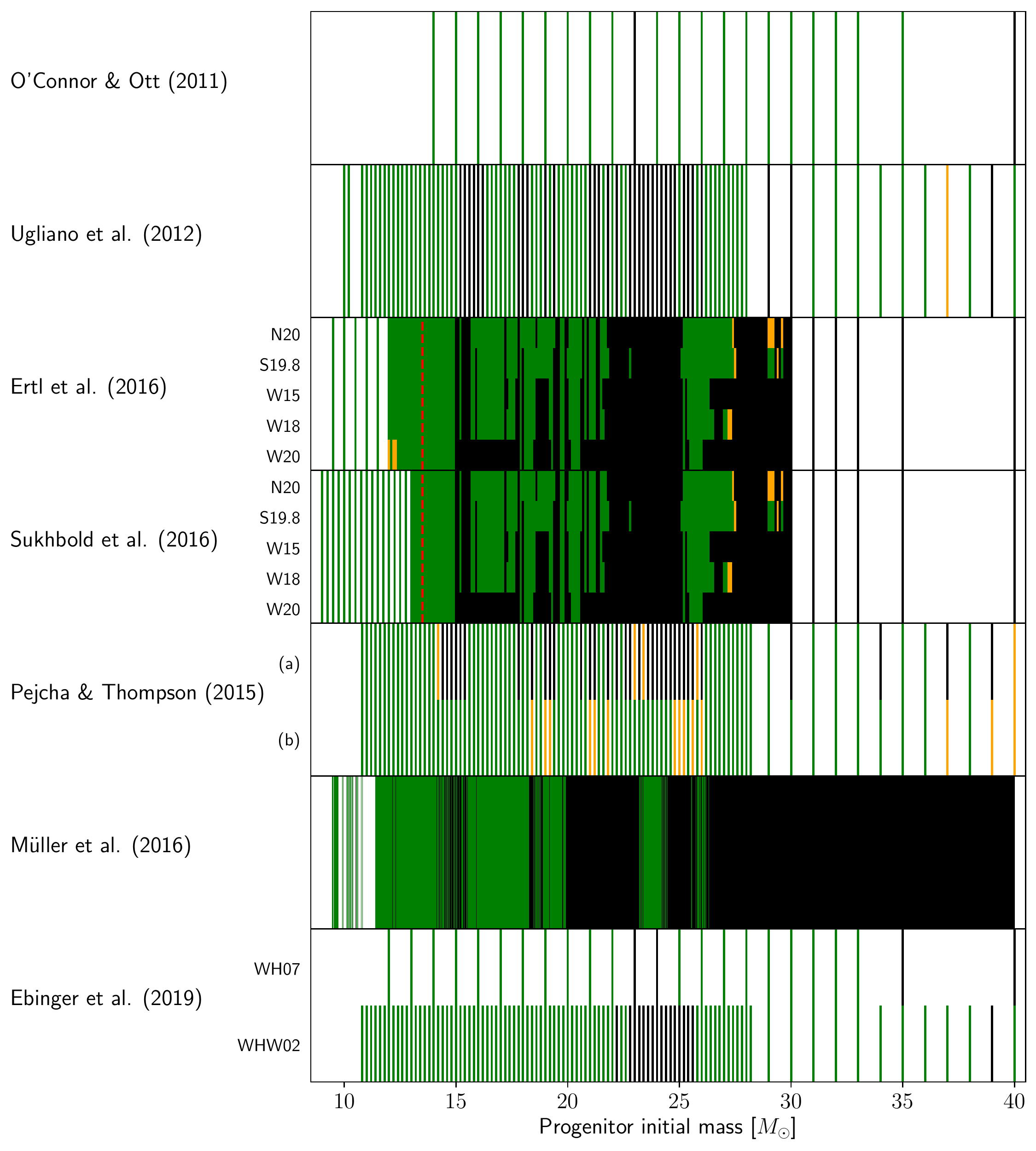}
    \caption{Outcome of the collapse of the core of a solar-metallicity massive star as a function of its initial mass. We show successful explosions leaving behind a neutron star (green), explosions with significant fallback (orange), and direct collapse to a black hole (black). Each bar corresponds to one progenitor model. To illustrate the differences between progenitor grids, the bar width is fixed at $0.1\,\msun$ except for \citet{muller16}, where it is set to $0.01\,\msun$. The results of \citet{oconnor11} are for \citet[WH07]{wh07} progenitor set and LS220 equation of state. \citet{ugliano12} used \citet[WHW02]{whw02} progenitors. \citet{ertl16} and \citet{sukhbold16} used a mixture of progenitors, which were exploded with five different calibrations of the supernovae engine (N20, S19.8, W15, W18, W20), as indicated in the Figure. The results are identical for progenitors above $13.5\,\msun$, which is marked with a vertical red dashed line. \citet{pejchathompson15} used two different parameterizations for WHW02 progenitors, where (a) has a fraction of non-exploding progenitors and (b) has explosions for all progenitors. The results of \citet{muller16} are for a custom grid of progenitors and their method cannot diagnose fallback explosions. The results of \citet{ebinger19} are for WHW02 and WH07 progenitor sets. }
    \label{fig:explodes}
\end{figure*}

\begin{figure}
    \centering
    \includegraphics[width=\textwidth]{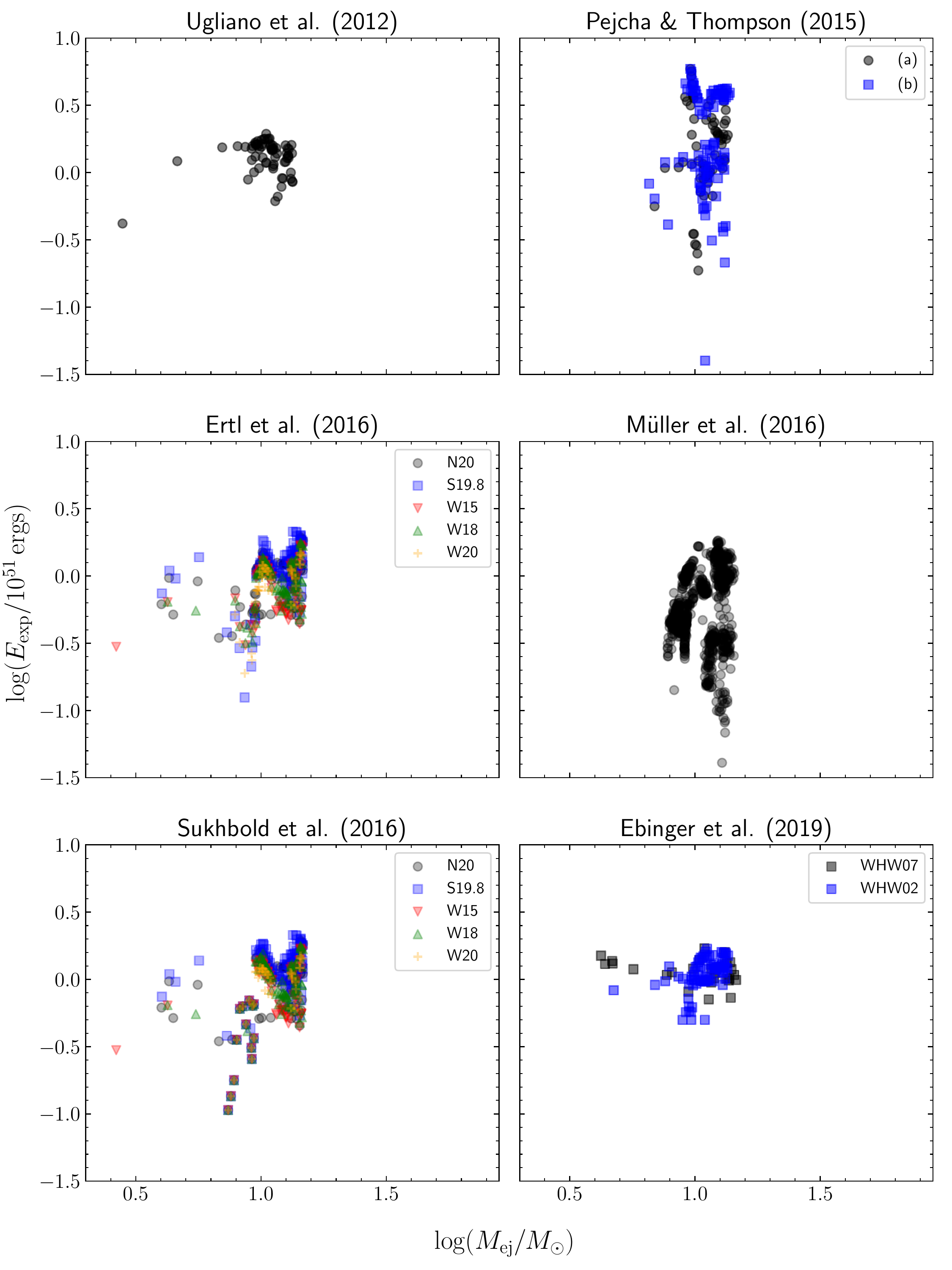}
    \caption{Explosion energy $\eexp$ as a function of ejecta mass $\mej$. The ejecta mass was determined as the final mass of the progenitor before the explosion with the remnant mass subtracted. Note that the final progenitor mass strongly  depends on very uncertain wind mass loss rates. This is different from quantities like $\eexp$ and $\mni$, which are set by the final core structure, which is not strongly affected by processes in the envelope. For WHW07 progenitors in \citet{ebinger19} we took the final mass from the progenitor set of \citet{ertl16}. Labels (a) and (b) refer to the two explosion parameterizations of \citet{pejchathompson15}. More detailed explanation of individual models is given in Figure~\ref{fig:explodes}.}
    \label{fig:mej_eexp}
\end{figure}

\begin{figure}
    \centering
    \includegraphics[width=\textwidth]{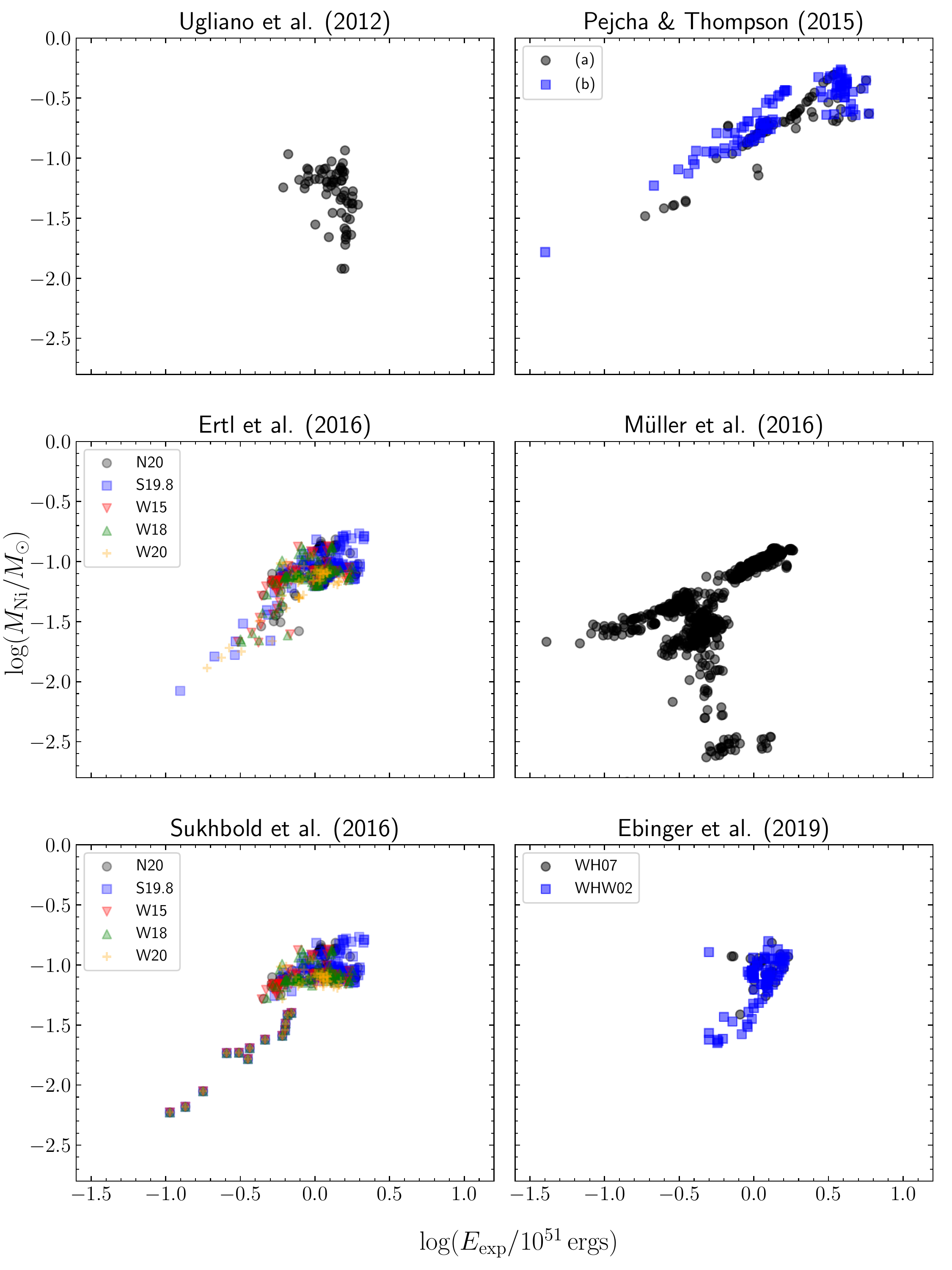}
    \caption{Mass of synthesized nickel as a function of explosion energy. For \citet{ertl16} and \citet{sukhbold16} we used the recommended value of the nickel plus half of the mass of tracer particles. Labels (a) and (b) refer to the two explosion parameterizations of \citet{pejchathompson15}.}
    \label{fig:eexp_mni}
\end{figure}

\begin{figure}
    \centering
    \includegraphics[width=\textwidth]{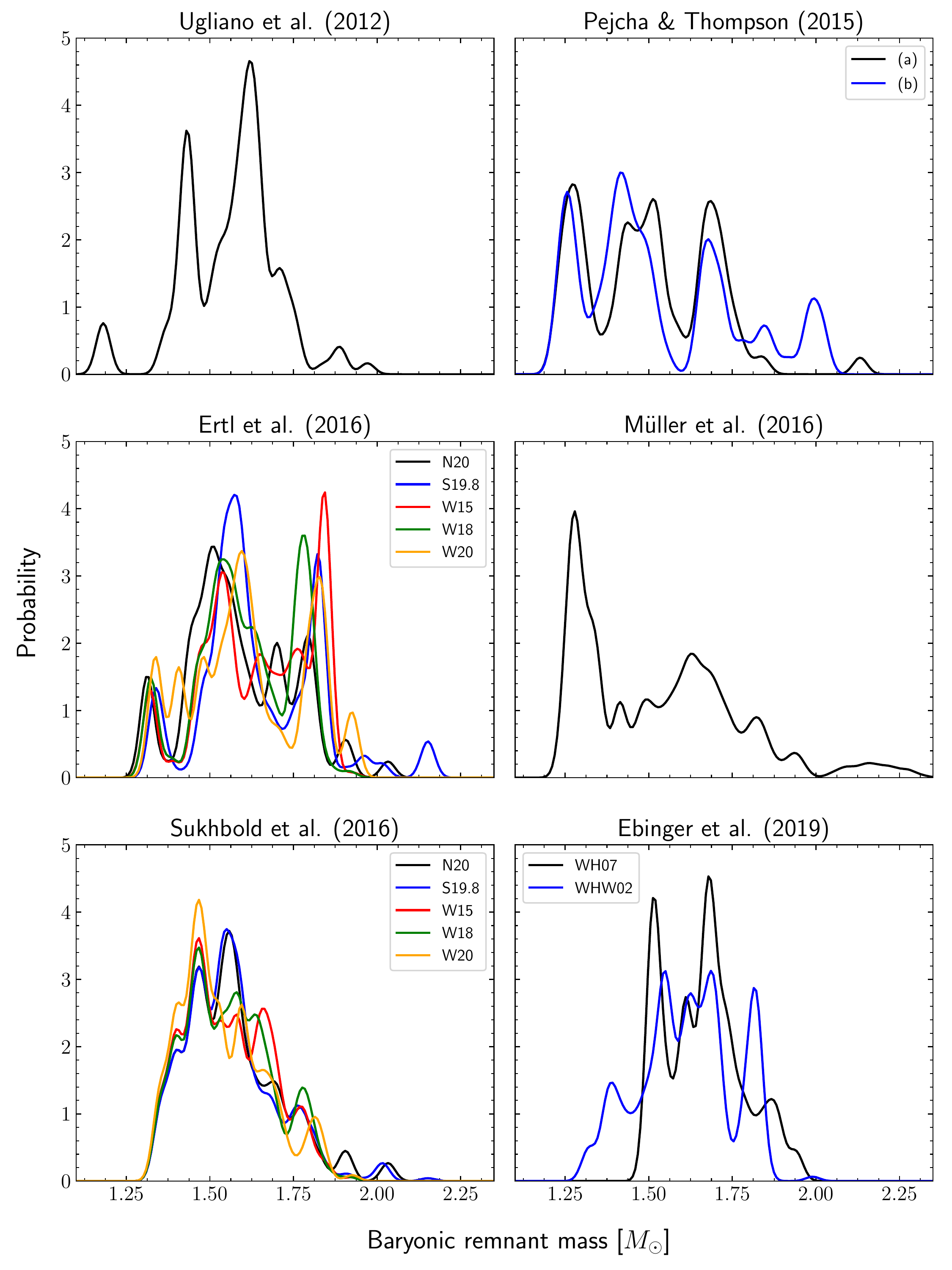}
    \caption{Probability distribution of baryonic masses of remnant neutron stars. The theoretical results are obtained by integrating over the provided progenitor grid, $M_i$ with weight given to $i$-th progenitor as $M_i^{-2.35} ( M_{i+1} - M_{i-1} )/2$, where the first factor is the \citet{salpeter55} initial mass function. The progenitor grid is assumed to be uniformly-spaced at its edges. The remnant mass distribution from each progenitor is described by a Gaussian with a width of $0.03\,\msun$. Labels (a) and (b) refer to the two explosion parameterizations of \citet{pejchathompson15}.}
    \label{fig:mrem}
\end{figure}

In this Section, we describe the theoretical efforts to quantify the progenitor--supernova connection for a large number of progenitor models. The results, when available, are summarized also graphically. Figure~\ref{fig:explodes} shows predictions of whether each progenitor explodes or not and whether the explosion exhibits substantial fallback. Figure~\ref{fig:mej_eexp} shows the predictions of explosion energy $\eexp$ as a function of ejecta mass $\mej$. Figure~\ref{fig:eexp_mni} displays the correlation between $\eexp$ and the mass of radioactive $^{56}$Ni synthesized in the supernova explosion, $\mni$. Finally, Figure~\ref{fig:mrem} provides predictions for mass distributions of neutron star remnants.

Can we predict the outcome of core collapse without running prohibitively large number of expensive multi-dimensional simulations? Unfortunately, supernova simulations in spherical symmetry consistently fail to explode except for the lowest-mass progenitors with tenuous envelopes. Still, spherical 1D calculations can reveal progenitor trends in the neutrino cooling of accreting material and how efficiently do the neutrinos heat the layers below the shock. In this spirit, \citet{oconnor11} defined the compactness of the pre-collapse progenitor
\begin{equation}
    \xi_M = \frac{M/\msun}{R(M_{\rm bary} = M)/1000\,{\rm km}},
    \label{eq:xi}
\end{equation}
which measures the concentration of mass at progenitor positions relevant for neutron star formation, $M\approx 1.2$ to $3\,\msun$. Higher $\xi_M$ implies the iron-core and its surrounding material is more centrally condensated, higher binding energies, and hence more difficult supernova explosions. Compactness changes during the collapse of the star so a common moment is needed to meaningfully compare different progenitors: \citet{oconnor11} argued that the moment of core bounce is a natural choice. \citet{oconnor11} also studied how artificially increasing the neutrino heating rate in their 1D general-relativistic simulations with neutrino leakage/heating scheme with code GR1D \citep{oconnor10} facilitates the explosion and found that progenitors with $\xi_{2.5} \lesssim 0.45$ are more likely to explode. They also noted that mapping between $\xi_M$ and the progenitor initial mass is non-monotonic. In most of the works summarized below, compactness is a relatively good predictor of the collapse outcome. 

\citet{ugliano12} used 1D Eulerian code with gray neutrino transport and replaced the neutron star core with an inner boundary condition parameterizing the  contraction and neutrino cooling of the proto-neutron star. Through a series of remappings the explosion was followed to $10^{15}$\,cm to determine $\eexp$ and to track fallback. Nuclear reaction network was used to estimate $\mni$. The free parameters of the neutron star cooling model were tuned to reproduce $\eexp$ and $\mni$ of SN1987A for the s19.8 red supergiant progenitor of \citet{whw02}. The same parameters were then utilized for the remaining about $100$ solar-metallicity progenitor models of the same series. This work showed for the first time that explosion properties can vary dramatically as a function of initial mass of the progenitor models. There are islands of non-explodability even at progenitor masses as low as $15\,\msun$. Interestingly, the scale of variations is comparable to the density of the progenitor grid of $0.1\,\msun$ pointing to deterministic chaos in the pre-supernova stellar evolution \citep{sukhbold14,sukhbold18}. \citet{ugliano12} do not predict a strong correlation between $\eexp$ and ejecta mass $\mej$ or $\mni$.

\citet{ertl16} revisited the model of \citet{ugliano12} with an updated equation of state, nuclear reaction network, larger set of progenitor models, and five different models of SN1987A progenitors chosen as calibrators: S19.8 from \citet{whw02}, W15 from \citet{woosley88}, W18 with rotation from unpublished results of Woosley, W20 from \citet{woosley97}, and N20 from \citet{shigeyama90}. Furthermore, the model for the excised core was modified for progenitors with initial masses $\le 13.5\,\msun$ so that these models exhibit weak explosions. The empirical support for this modifications is SN1054 with estimated progenitor mass of $10\,\msun$ and explosion energy of only $10^{50}$\,ergs \citep{smith13,yang15}. \citet{ertl16} found that the combination of mass coordinate and its derivative at a location of entropy per baryon equal to 4 outperforms the compactness in predicting the outcome of the collapse.

\citet{sukhbold16} used essentially the same basic setup as \citet{ertl16}, but with updated progenitor models below $13.5\,\msun$ and different explosion calibration in this mass range (red dashed line in Fig.~\ref{fig:explodes}). They also post-processed the results with a hydrodynamics code KEPLER to provide nucleosynthetic yields for up to 2000 nuclei and light curves of the supernovae. Recently, \citet{ertl19} further improved their method and applied it to helium progenitors evolved with mass loss, which should approximate effects of binary star evolution. They find increased fraction of fallback explosions leading to small population of remnants in the ``mass gap'' between $2$ and $5\,\msun$.

\citet{pejchathompson15} evolved the accretion phase with GR1D code \citep{oconnor10} for several seconds to obtain consistent runs of proto-neutron star mass, radius, and neutrino luminosity and energy. Based on these trajectories, they estimated the  time-evolution of the critical neutrino luminosity under the assumption of quasi steady-state evolution. The actual neutrino luminosities never crossed the critical value so \citet{pejchathompson15} parametrically modified the critical neutrino luminosity and consistently applied the changes to the progenitor suite of \citet{whw02}. They estimated $\eexp$ as a time-integrated power of neutrino-driven wind and $\mni$ as a mass in volume exposed to sufficiently high temperatures. Their results are qualitatively similar to other works from parameterized explosions, with \citet{pejchathompson15} showing that the pattern does not dramatically change with different choices of the parameterization.

\citet{muller16} constructed a model of accretion phase evolution using only ordinary differential equations, which heuristically accounts for simultaneous accretion and outflows. With only little computing time required, \citet{muller16} applied the method to over 2000 progenitor models and, similarly to earlier results, they estimated remnant mass, $\eexp$, and $\mni$. For the correlations visualized here, the results are in qualitative agreement with other works except for relatively small population of explosions with very small $\mni$ and normal $\eexp$.

\citet{ebinger19} presented results of artificial explosions based on the PUSH method. This setup was initially implemented by \citet{perego15} and utilizes a general relativistic hydrodynamics code in spherical symmetry \citep{liebendorfer01} with an isotropic diffusion source approximation for the transport of electron neutrinos and antineutrinos. Parameterized explosions are triggered by introducing a heating term proportional to the luminosity of $\mu$ and $\tau$ neutrinos, which normally affect the evolution less than electron neutrinos and antineutrinos. The magnitude of the heating is scaled by the progenitor compactness. \citet{ebinger19} provide the basic explosion predictions for solar-metallicity progenitors, while followup work gives predictions of detailed nucleosynthesis \citep{curtis19} and for low-metallicity progenitors \citep{ebinger19b}. The results of this approach are in broad agreement with other results except that the ``bar-code'' pattern in the explodability is not as prominent. However, other parameters such as $\eexp$ are also not monotonic with the progenitor initial mass.

The predictions of parameterized explosions have been continuously evolving by including additional physics. \citet{couch19} added modified mixing length theory of convection and turbulence to spherically-symmetric hydrodynamical equations and studied explodability of 138 solar-metallicity progenitors as a function of the mixing length parameterization. Similarly to most other works, they find complex landscape of explosions and failures, and provide predictions of explosion energies and remnant masses. A similar approach was implemented by \citet{mabanta19} using the turbulence model of \citet{mabanta18}. The appropriate form of the turbulence model is still debated \citep{muller19}. \citet{nakamura15} performed 2D simulations of neutrino-driven explosions of \citet{whw02} progenitors and found that most of them are exploding. The explosion properties like $\eexp$, $\mni$, and neutron star mass correlate with the compactness.

It is worth noting that many of the parameterized explosion schemes and explicitly tied to the observed properties of SN1987A and SN1054. SN1987A was a peculiar explosion from a blue supergiant progenitor, which might have experienced a merger shortly before the supernova explosion \citep[e.g.][]{chevalier89,hsn_podsiadlowski}. Kinetic energy in the Crab remnant is low, which contrasts with the apparently normal light curve \citep[e.g.][]{hester08,smith13}. It is unclear how are the theoretical predictions affected by peculiarities of the calibration source. The observed properties of calibrators are never exactly matched by the theoretical models and the observations always have a margin of uncertainty. In a sense, the calibration models serve as a zero-point for the entire population. If the calibration model properties were vastly different from a typical progenitor with roughly the same mass, the supernova \emph{population} would be discrepant in quantities like total amount of synthesized $\nickel$, the results supernovae would be of too high or too low luminosity when compared to observations, etc. As we will illustrate below, this does not seem to be the case with the currently available observations and theoretical predictions. Still, some of these uncertainties might be absorbed in internal tunable model parameters.

Finally, we emphasize that parameterized models predominantly explore physical effects that were explicitly included in their construction and that the possibility of discovering new effects or their combinations is narrower than what is possible in more-complete multi-dimensional simulations with less prior assumptions. Nonetheless, one way to declare that features of the most complicated simulation are fully understood is that it is possible to replicate these results within some kind of parameterized model.

\section{Observational efforts}

\begin{figure}
    \centering
    \includegraphics[width=\textwidth]{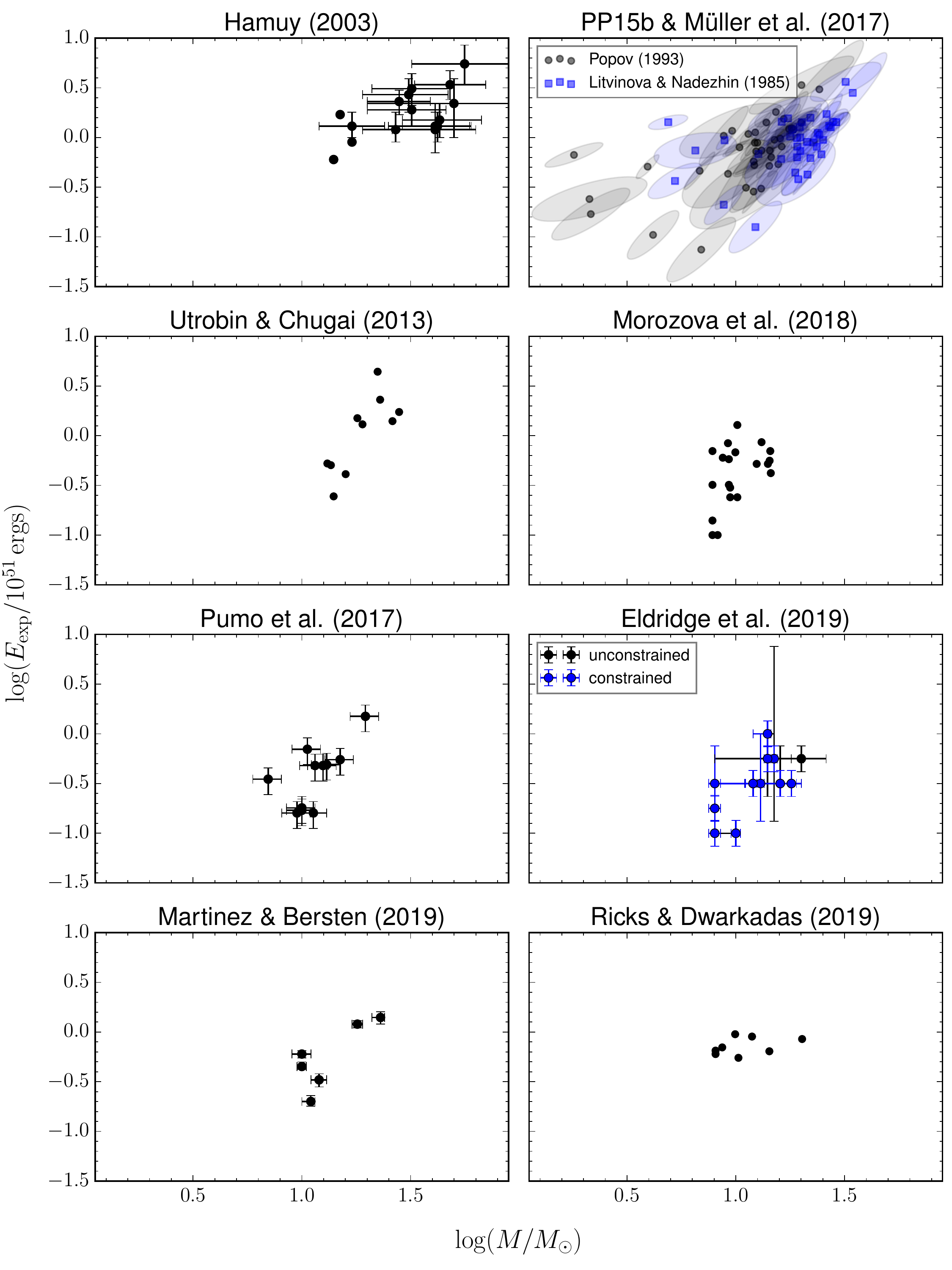}
    \caption{Explosion energy $\eexp$ as a function of ejecta mass $\mej$ or progenitor initial mass $M$, as inferred from modeling of non-interacting hydrogen-rich (mostly Type II-P) supernova light curves and expansion velocities. Works based on analytic scaling relations \citep{hamuy03,pejcha15b,muller17} report values for $M$, which are related to ejecta mass. \citet{utrobin19}, \citet{pumo17}, and \citet{morozova18} report ejected envelope mass. \citet{martinez19} report stellar mass just prior to the explosion. \citet{eldridge19} report progenitor initial mass. The caveat of Fig.~\ref{fig:mej_eexp} concerning the uncertain wind mass loss rates affecting the inferences of initial progenitor mass applies here as well.}
    \label{fig:mej_eexp_obs}
\end{figure}

\begin{figure}
    \centering
    \includegraphics[width=\textwidth]{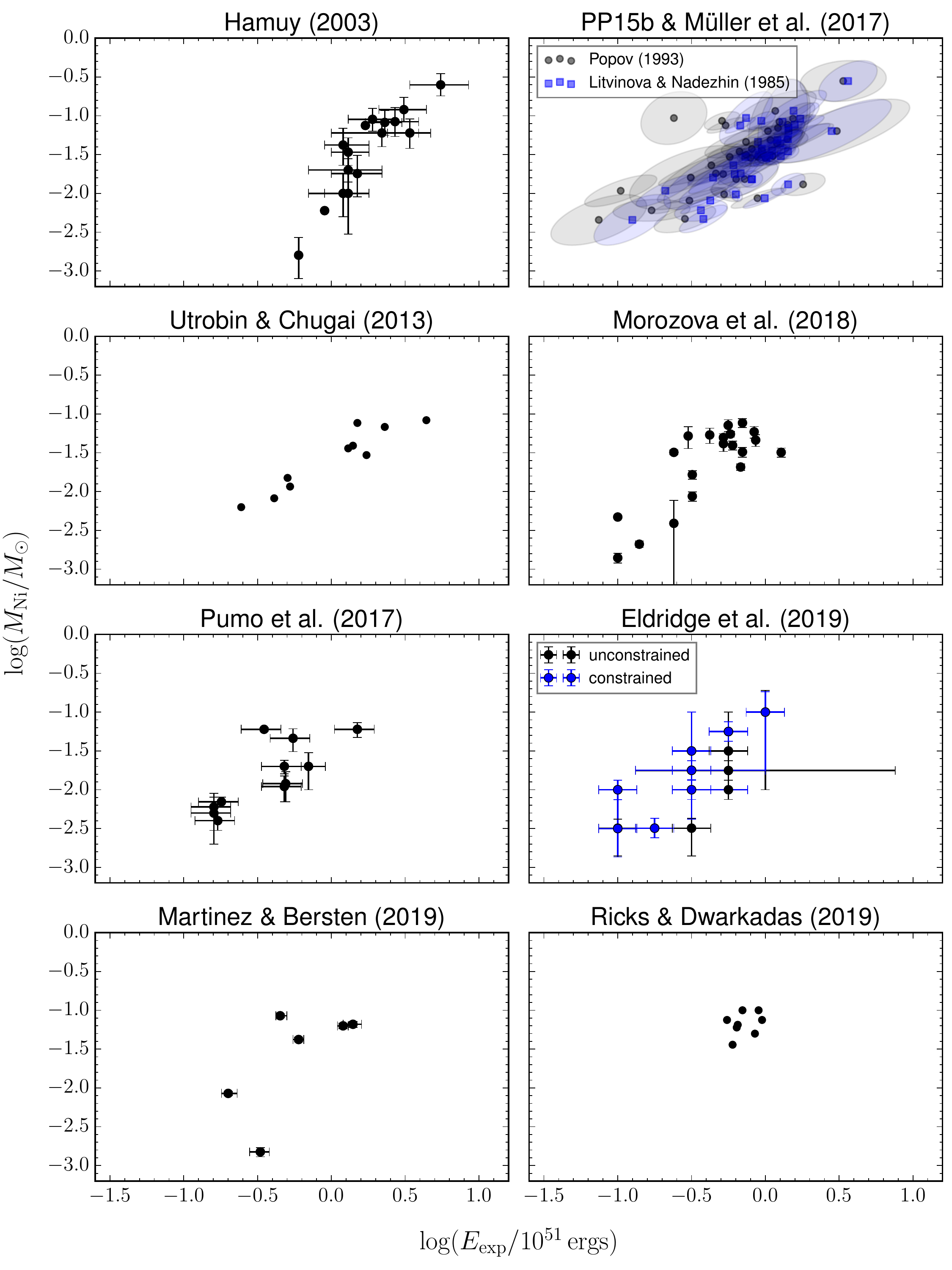}
    \caption{Nickel mass $\mni$ and explosion energy $\eexp$ from modeling of non-interacting hydrogen-rich (mostly Type II-P) supernovae light curves.}
    \label{fig:eexp_mni_obs}
\end{figure}

\begin{figure}
    \centering
    \includegraphics[width=0.6\textwidth]{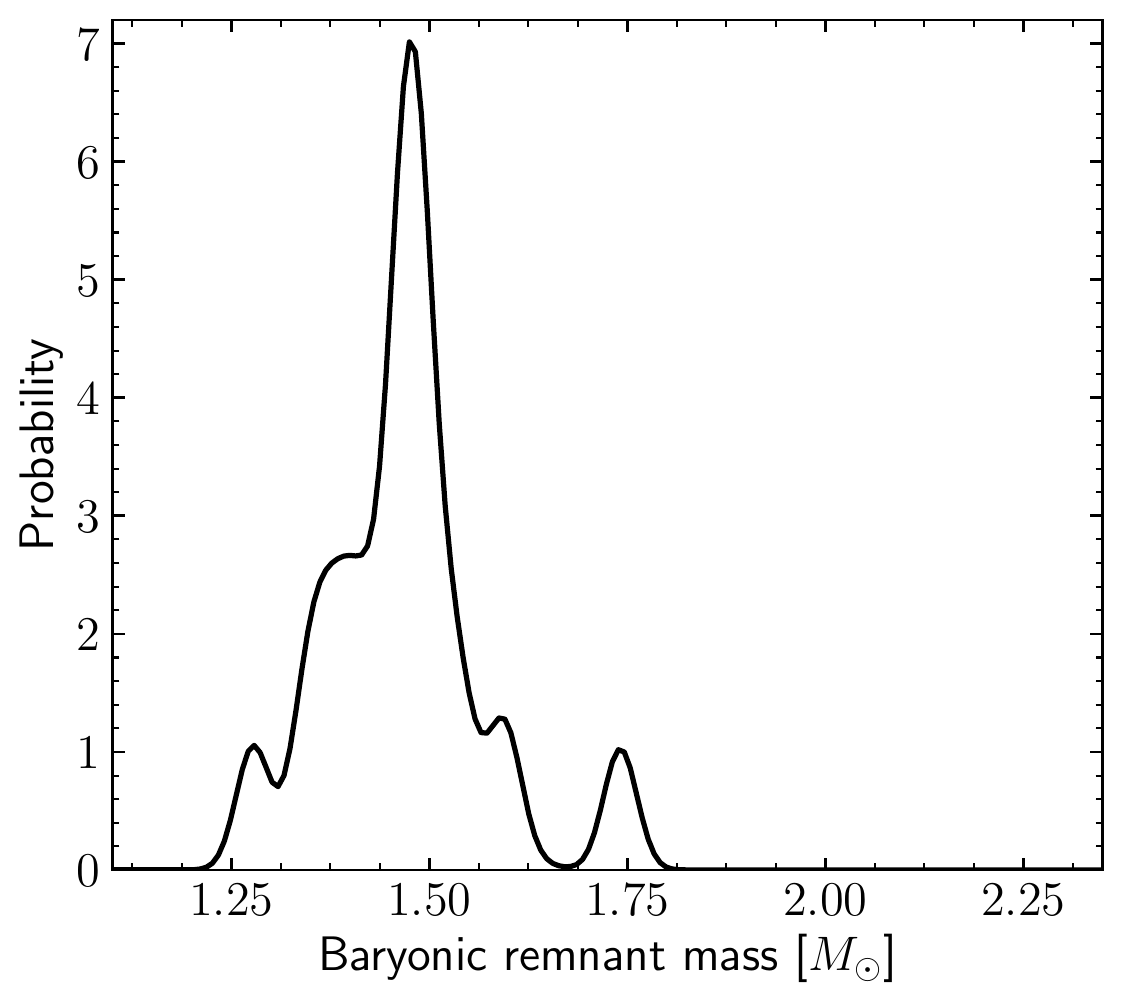}
    \caption{Distribution of bayronic masses of double neutron stars and non-recycled pulsars from \citet{ozel16}. Each measurement is approximated with a Gaussian with a width of the quoted uncertainty value or $0.03\,\msun$, whichever is larger. The values were converted from observed gravtiational masses to baryonic masses using the equation of \citet{timmes96}. This figure should be compared with the theoretical predictions depicted in Fig.~\ref{fig:mrem}.}
    \label{fig:nsmass}
\end{figure}

Theory predicts discrete islands of successful and failed explosions within a population of massive stars. A straightforward way to observationally verify these predictions is to witness a massive star collapsing to a black hole without an accompanying supernova. This avenue has been pursued with both ground-based \citep{kochanek08,adams17b} and space-based data \citep{reynolds15} and have finally yielded a single candidate \citep{gerke15,adams17} with a mass of $\sim 25\,\msun$. In a similar spirit, archival pre-explosion images of supernova positions can be used to infer distribution of progenitor masses of known supernovae. These observations have suggested an unexpected lack of red supergiant progenitors with initial masses $\gtrsim 17\,\msun$ \citep{kochanek08,smartt09,smartt15}. Both the disappearing star and progenitor non-detections are naturally explained by a lack of explosions in this mass range seen in parameterized models (Fig.~\ref{fig:explodes}). Masses of exploding stars can be also estimated from the age of the surrounding population, although the precision is lower than for directly imaged progenitors \citep{badenes09,anderson12,jennings12,williams19}, likely too rough to resolve the complex landscape of supernova explosions.

Observational studies relying on direct imaging are only feasible in a relatively small volume with correspondingly low rate of supernova explosions. But supernovae themselves are visible from great distances and their light could provide complementary observational evidence. The luminosity, duration, and shape of the supernova light curve depend on the progenitor envelope mass, size, structure, and composition, the explosion energy, and the amount of radioactive nuclei and how they are mixed within the envelope. Hereafter, we narrow the focus on the relatively common Type II-P supernovae, which are explosions of red supergiant progenitors with hydrogen envelopes. Type II-P supernovae are useful as theoretical benchmarks, because the progenitor structure should closely match the result of evolution of single stars. This does not mean that binary interactions did not occur in the previous evolution, in fact, a fraction of Type II-P supernovae are possibly the result of stellar mergers or other mass exchanges in binary stars \citep{zapartas19}.

Light curves of Type II-P supernova in the first $100$ days or so are dominated by the diffusion of light out of a recombining hydrogen envelope, which leads to a ``plateau'' of relatively constant bolometric luminosity. The dramatic change of opacity at hydrogen ionization implies that the supernova has a relatively well-defined photosphere, which moves inward in the Lagrangian mass coordinate of the ejecta. Simple analytical estimates show that the three principle observables --- plateau luminosity, duration, and spectroscopic expansion velocity --- are determined by three intrinsic characteristics --- progenitor radius, ejecta mass, and supernova explosion energy. Therefore, it is tempting to turn the observables into quantities interesting for the theory of the explosion mechanism \citep[e.g.][]{popov93,kasen09,sukhbold16}.

After the hydrogen envelope fully recombines, the luminosity drops and starts to closely track energy input from the decay of radioactive elements synthesized in the explosion. Early on, the most important radioactive chain starts with $^{56}$Ni, although the actual decaying element powering the radioactive tail is its decay product $^{56}Co$. The initial mass of $^{56}Ni$ is thus the relevant quantity to study.  Assuming full trapping of the radioactive decay products, the normalization of the exponentially decaying light curve yields an estimate of $\mni$. The temperatures required for synthesis of $^{56}$Ni imply that its formation occurred within few thousand km of the center of the supernova explosion, which makes $\mni$ a useful probe of the explosion development and internal structure of the progenitors.

Here, we summarize a subset of efforts to estimate physical parameters out of Type II-P supernova light curves and spectra. There is a great range of techniques ranging from simple scaling relations to full radiation hydrodynamical models and spectrum fitting. Figure~\ref{fig:mej_eexp_obs} shows $\eexp$ as a function of ejecta or progenitor mass. The two masses should differ by $1$--$2\,\msun$ for explosions leaving behind a neutron star. Figure~\ref{fig:eexp_mni_obs} explores the correlation between $\eexp$ and $\mni$. These Figures should be directly compared with theoretical results in Figures~\ref{fig:mej_eexp} and \ref{fig:eexp_mni}.

The simplest approach to estimating explosion parameters relies on combining consistently inferred basic properties of the light curves and velocities with analytic scaling relations.
\begin{itemize}
\item \citet{hamuy03} used analytic scaling relations of \citet{litvinova85} to estimate physical parameters of hydrogen-rich supernovae. He found that these supernovae span a range of physical parameters, and characterized correlations between ejecta mass, $\eexp$, and $\mni$ in the sense that more massive stars produce more energetic explosions and more $^{56}$Ni.
\item \citet{pejcha15a} constructed a global hierarchical model of multi-band light curves and expansion velocities and trained it on a sample of nearby well-observed supernovae. Using the global covariance matrix of the fit, \citet{pejcha15b} inferred explosion properties using \citet{litvinova85} and \citet{popov93} scaling relations taking into account all uncertainties in the model. \citet{muller17} nearly doubled the sample with the same method.
\end{itemize}

A number of groups used radiative hydrodynamics codes to model observed supernova light curves. In most cases, these are Lagrangian, spherically-symmetric codes with one group radiation transfer. Often, flux-limited diffusion is utilized to connect optically-thick and optically-thin regions. Since the number of supernovae with densely-covered light curves and frequent spectroscopic observations is limited, the supernova samples often overlap. 
\begin{itemize}
    \item \citet{utrobin19} summarized physical parameters of Type II-P supernovae from their previous work \citep[e.g.][]{utrobin13,utrobin15,utrobin17} with radiation hydrodynamics code that solves separately for the temperatures of gas and radiation \citep{utrobin04}. 
    \item \citet{pumo17} provided an overview of their previous work \citep[e.g.][]{spiro14,takats14,takats15,huang15} using a general-relativistic hydrodynamics code, which solves the first two moments of the radiative transfer equations \citep{pumo11}. 
    \item \citet{martinez19} report fits with a code assuming flux-limited diffusion and local thermodynamic equilibrium for the gas and radiation \citep{bersten11} for supernovae with independent progenitor mass estimates from pre-explosion imaging. 
    \item A similar code was developed and made publicly available as SNEC by \citet{morozova15}, which was followed by fits to light curves by \citet{morozova18}, also taking into account interactions of the supernova blast wave with the circumstellar medium in the early parts of the light curves. They also presented the $\chi^2$ surfaces of their fits. 
    \item \citet{eldridge19} compared their database of supernova light curves calculated with SNEC and binary population synthesis stellar models \citep{eldridge18} to infer explosion parameters of a sample of well-observed supernovae with pre-explosion progenitor detections. 
    \item \citet{ricks19} used MESA stellar evolution code to simulate supernova progenitors, explode them, and calculate the resulting optical light curves with multi-group radiative hydro code STELLA \citep{paxton15,paxton18,blinnikov98,blinnikov06}. The dynamical range of the inferred parameters is more limited than in other works, which is probably due to relatively small sample size.
\end{itemize}

Finally, Figure~\ref{fig:nsmass} shows the observed distribution of a subset of neutron star masses, which should be close to the birth mass. This includes double neutron star binaries and non-recycled pulsars \citep{ozel16}. The observations should be compared with theoretical results in Figure~\ref{fig:mrem}. A more quantitative comparison with theoretical models was done by \citet{pejcha_ns} and \citet{raithel18}. The latter work finds that single-star models have trouble explaining high-mass neutron stars. They also looked at the distribution of black hole masses.

\section{How to compare observations with theory}

We see that there is a general qualitative agreement between the theoretical and observational results for normal hydrogen-rich supernovae: most theories predict explosion energies of $10^{50}$ to $10^{51}$ ergs, nickel masses of $0.01$ to $0.1\,\msun$, and baryonic neutron star masses between $1.3$ and $1.7\,\msun$, as is observed. This could be viewed as a success of the theory, but is not entirely unexpected. All models have some free parameters, which can be tuned to achieve the desired outcome. An example is relating one of the free theoretical parameters to the progenitor compactness. To first order, these choices affect more significantly quantities like population means or medians and less characteristics like correlations between different explosion outcomes, their slopes, and intrinsic scatters. 

Observational inferences are not free from similar biases either. Analytic scaling relations always include an absolute term, which can be uncertain. As a result, works based on these relations have a considerable freedom in rescaling all of the physical quantities by a factor. Relative positions of individual supernovae are affected less. The uncertainties in fitting based on radiation hydrodynamics are more intricate, but have led to overestimates of progenitor masses when compared to inferences from pre-explosion images \citep[e.g.][]{utrobin09,maguire10}. Recent works, however, indicate a better agreement between the two independent methods \citep{martinez19}. 

Recently, an additional challenge has been recognized with inferences of explosion parameters from the light curve plateaus of Type II-P supernovae: estimates of explosion energy and ejecta mass are nearly degenerate even in radiation hydrodynamical models. Qualitatively, the degeneracy arises because higher ejecta mass leads to a longer diffusion time, which can be compensated for with higher explosion energy. If explosion energy per unit ejecta mass remains nearly constant, the observed expansion velocity, and the plateau luminosity and duration do not change much \citep{arnett80}. \citet{nagy14} found explosion parameter correlations with a semianalytical code. \citet{pejcha15b} argued that analytic scaling relations point to nearly degenerate inferences of explosion energy and ejecta mass so that uncertainties in  quantities like distance and reddening will manifest as a spread along a diagonal line in $\eexp$--$\mej$ plot. They argued that this might be responsible for the claimed observational correlation between these two quantities. \citet{goldberg19} used MESA and STELLA codes to illustrate that additional information is needed to break the degeneracy between $\eexp$, $\mej$, and the progenitor radius (see especially their Fig.~26). They also argued that spectroscopic velocities secured during the shock-cooling phase in the first $\sim 15$ days after the explosion could break the degeneracy. Similar conclusions were independently reached also by \citet{dessart19} with multi-group radiation transfer code, which models full spectra.

The possibility of breaking the degeneracies with early observations suggested by \citet{goldberg19} might not work for all Type II-P supernovae. There has been growing amount of evidence that some red supergiant progenitors are surrounded by circumstellar medium very close to the surface, which influences the early part of the light curve and spectra. The evidence includes modeling of early light curves \citep{morozova15,morozova17,morozova18,moriya17,moriya18,dessart17,forster18} and ``flash spectroscopy'' of the progenitor surroundings \citep[e.g.][]{khazov16,yaron17,hosseinzadeh18}.

Although a significant fraction of the population of stripped-envelope supernovae (spectroscopic types IIb, Ib, and Ic) likely originates from binary interactions and their ejecta masses are not representative of the progenitor initial stellar mass, explosion energies and nickel masses inferred from light curve modeling can test the explosion theories as well. Recently, \citet{ertl19} applied the parameterized  neutrino mechanism to a population of helium stars of various masses. They found that the observed value of $\mni$ are noticeably higher than the allowed range of theoretical predictions. Perhaps additional sources of energy from (magneto)-rotational processes can provide the necessary boost. It is then tempting to speculate whether the neutrino mechanism is subdominant even in normal supernovae \citep{sukhbold17}.

The limitations of both theory and observations imply that moving forward with deeper and more quantitative test of supernova explosion mechanism will require investigating finer details in the observationally inferred explosion properties. There are several possibilities to move forward.  

The degeneracies in inferring the explosion properties during the plateau can be put aside by focusing on the later phases dominated by radioactive decay of $^{56}$Ni. This is the approach taken by \citet{muller17} and \citet{anderson19}. However, Type II-P supernovae are faint during the radioactive decay phase and the inferences of $\mni$ depend on knowing well the explosion date and the distance. Future time-domain surveys could help with some of these challenges.

The alternative is to take the degeneracies explicitly into account when making the inferences and when doing comparisons with the theory. While degeneracies can be easily quantified with the covariance matrix or by directly exploring the likelihood space with Markov Chain Monte Carlo techniques, it is less straightforward how to construct the correct model. The theoretical light and velocity curves from radiation hydrodynamics never match the observations within the level of their uncertainties. As a result, the sampling of observations will then bias the inferences. Furthermore, taking into account circumstellar interaction increases the number of free parameters, where some of them might be degenerate with the explosion properties. Some of these issues might be alleviated by introducing a ``transfer function'' between the theoretical light and velocity curves and the observed magnitudes and velocities. This approach requires a sufficiently large training dataset to constrain the large number of additional parameters. Finally, comparisons of supernova explosion properties with theory usually implicitly assume that the observations are representative of the underlying stellar populations. Taking into account selection effects might yield new tests of the explosion mechanism, for example, by comparing relative rates of low- and high-$\mni$ events.

\begin{acknowledgement}
I am grateful to Joseph Anderson and Tuguldur Sukhbold for insightful and useful referee reports. I thank Thomas Ertl, Thomas Janka, and Bernhard M\"{u}ller for sharing machine-readable versions of their results. This work has been supported by Primus grant PRIMUS/SCI/17 from Charles University, Horizon 2020 ERC Starting Grant ``Cat-In-hAT'' (grant agreement \#803158) and INTER-EXCELLENCE grant LTAUSA18093 from the Czech Ministry of Education, Youth, and Sports.
\end{acknowledgement}

\bibliographystyle{kluwer}
\bibliography{references.bib}

\end{document}